\documentclass[namedreferences]{SolarPhysics}
\usepackage[optionalrh]{spr-sola-addons} 
\usepackage{epsfig}                     
\usepackage{graphicx}                    
\usepackage{color}                       
\usepackage{url}                         
\usepackage{rotating}

\begin{document}

\begin{article}

\begin{opening}

\title{Accuracy and Limitations of Fitting and Stereoscopic Methods to Determine the Direction of Coronal Mass Ejections from Heliospheric Imagers Observations}

%
\author{N.~\surname{Lugaz}}

%
\runningauthor{Lugaz}
\runningtitle{Methods for Determining CME Directions}

%
  \institute{ Institute for Astronomy - University of Hawaii-Manoa, 2680 Woodlawn Dr., Honolulu, HI 96822
                     email: \url{nlugaz@ifa.hawaii.edu}}

\begin{abstract}
Using data from the Heliospheric Imagers (HIs) onboard STEREO, it is possible to derive the direction of propagation of coronal mass ejections (CMEs) in addition to their speed with a variety of methods. For CMEs observed by both STEREO spacecraft, it is possible to derive their direction using simultaneous observations from the twin spacecraft and also, using observations from only one spacecraft with fitting methods. This makes it possible to test and compare different analyses techniques.
In this article, we propose a new fitting method based on observations from one spacecraft, which we compare to the commonly used fitting method of \inlinecite{Sheeley:1999}. We also compare the results from these two fitting methods with those from two stereoscopic methods, focusing on 12 CMEs observed simultaneously by the two STEREO spacecraft in 2008 and 2009. We find evidence that the fitting method of \inlinecite{Sheeley:1999} can result in significant errors in the determination of the CME direction when the CME propagates outside of $60^\circ \pm 20^\circ$ from the Sun-spacecraft line.  We expect our new fitting method to be better adapted to the analysis of halo or limb CMEs with respect to the observing spacecraft. We also find some evidence that direct triangulation in the HI fields-of-view should only be applied to CMEs propagating approximatively towards Earth ($\pm 20^\circ$ from the Sun-Earth line). Last, we address one of the possible sources of errors of fitting methods: the assumption of radial propagation. Using stereoscopic methods, we find that at least seven of the 12 studied CMEs had an heliospheric deflection of less than 20$^\circ$ as they propagated in the HI fields-of-view, which, we believe, validates this approximation. 

\end{abstract}
\end{opening}

%
%

\section{Introduction} \label{intro}

Thousands of coronal mass ejections (CMEs) have been observed remotely by coronagraphs and hundreds by {\it in situ} instruments since the 1970s (for a review of CME observations, see, for example \opencite{Hundhausen:1993} and \opencite{Howard:2006}). 
In the past five years, with the launch of spacecraft carrying heliospheric imagers ({\it Coriolis} and the {\it Solar-Terrestrial Relations Observatory} (STEREO)), we have witnessed the start of a new era, where CMEs can be routinely observed to radial distances as far as 0.5 AU with the Heliospheric Imagers (HIs, see \opencite{Eyles:2009}) and sometimes up to Earth's orbit. Their remote properties can now be compared to {\it in situ} measurements \cite{Davis:2009,Harrison:2009, Moestl:2009b}. To analyze these new measurements, new methods and approximations must be devised. Shortly after the launch of {\it Coriolis} and STEREO, two simple ways to analyze wide angle heliospheric observations were developed: the Point-P \cite{Vourlidas:2006} and Fixed-$\Phi$ \cite{Kahler:2005} approximations. Recently, another approximation has also been proposed (harmonic mean (HM), see \opencite{Lugaz:2009c}) and these approximations have been completed by other techniques: visual fitting \cite{Wood:2009a,Maloney:2009}, fitting to a family of pre-existing simulated ejections \cite{THoward:2009b, THoward:2009a} and fitting to known functions of the speed and direction \cite{Rouillard:2008}. These techniques can derive, in addition to the CME speed and position, its average direction of propagation. Two other methods have been proposed to analyze simultaneous (i.e. stereoscopic) CME observations in the heliosphere, by direct triangulation \cite{Liu:2010} or by a ``tangent to a sphere'' method \cite{Lugaz:2010b}, to obtain, at all times, the CME position and direction of propagation. 

Heliospheric instruments are planned in a number of future missions, including the {\it Solar Orbiter} and the {\it Solar Probe}. However, it is possible that there will not be other stereoscopic observations of CMEs by heliospheric imagers after the end of the STEREO mission. The time period from early 2008, when the STEREO spacecraft separation reached 45$^\circ$ to the early 2010 when it reached about $130^\circ$ is the optimal period to have stereoscopic heliospheric observations. It should be used to validate, test and compare methods to analyze white-light heliospheric observations of CMEs. Such comparisons between methods was recently performed by \inlinecite{Davis:2010} between the visual fitting of COR images by \inlinecite{Thernisien:2009} and the analytical fitting to a constant direction and velocity by \inlinecite{Sheeley:1999}. The theoretical error and bias associated with the manual selection of the elongation data for the method of \inlinecite{Sheeley:1999} were also recently quantified in \inlinecite{Williams:2009}. In the present article, we compare two different stereoscopic heliospheric methods with each other and with fitting methods to study the direction of propagation of CMEs.

In this article,  in Section \ref{mono}, we propose a new fitting method, similar to that of  \inlinecite{Sheeley:1999} but based on the model of \inlinecite{Lugaz:2009c}. We compare theoretically the two methods based on stereoscopic observations proposed by \inlinecite{Liu:2010} and \inlinecite{Lugaz:2010b} in Section \ref{stereo}. In Section \ref{comp}, we analyse in detail one of the 12 CMEs observed simultaneously by the two STEREO spacecraft in 2008 and 2009, before doing a statistical comparison of the methods based on these real analyses. The conclusions of this investigation are drawn in Section \ref{conclusions}.

\section{CME Direction of Propagation from Single-Spacecraft Observations} \label{mono}

In the field-of-view of an heliospheric imager, the simple assumption of ``plane-of-the-sky'' cannot be made to derive the position of a density structure. The position is therefore measured as the angle between the observing spacecraft, the Sun and the density structure, and it is commonly referred as the elongation angle.
In the simplest case of a single plasma element or a spherical ejection, the elongation angle is a complex function of the heliocentric distance and of the angle of propagation with respect to the observing spacecraft. Therefore, the shape of the elongation vs. time curve depends on the speed and direction of propagation of a transient \cite{Sheeley:1999}. When CMEs are observed to large elongation angles (up to 40$^\circ$ and beyond), the elongation vs. time profile can be fitted to analytical functions and the average speed and average direction of the CME can be derived under certain assumptions (see, for example, \opencite{Webb:2009} and \opencite{THoward:2009a}). 

In the rest of this article, we note $\alpha$ as the elongation angle, and $\beta$ as the direction of propagation of the CME, following the terminology of \inlinecite{Rouillard:2008}. We also note $d_\mathrm{ST}$ as the heliocentric position of the observing spacecraft (here STEREO) and $t$ as the time.
 
\subsection{Methods}

We assume that an heliospheric imager observes a single plasma element, and we further assume that the direction of propagation is fixed. Under these assumptions, the heliocentric distance, $R$, can be derived analytically as a function of the elongation angle: 
$$
R_{\mathrm{F}\Phi} = d_\mathrm{ST}\frac{\sin \alpha}{\sin(\alpha + \beta_{\mathrm{F}\Phi})}.
$$
This relation is usually referred to as the Fixed-$\Phi$ (F$\Phi$) approximation (e.g., \opencite{Kahler:2005}).
It can be inverted \cite{Sheeley:1999}, assuming a constant velocity (CV), $V = R t$, resulting in:
\begin{equation}
\alpha = \arctan \left (\frac{ V_{\mathrm{F}\Phi CV}  t \sin\beta_{\mathrm{F}\Phi CV} }{d_\mathrm{ST} - V_{\mathrm{F}\Phi CV} t \cos\beta_{\mathrm{F}\Phi CV}} \right).\label{eq:TEFF}
\end{equation}
 
A measured profile of elongation vs. time can be fitted to a profile of calculated elongations given by Equation~(1). The Fixed-$\Phi$ constant velocity (F$\Phi$CV) procedure gives the average speed and direction of the transient. It was originally proposed by \inlinecite{Sheeley:1999} for slow ejections in the LASCO coronagraph fields-of-view. \inlinecite{Rouillard:2008} and \inlinecite{Sheeley:2008a} applied this technique to corotating interaction regions observed by the STEREO/HIs. It has since been widely used to study CMEs (e.g., in \opencite{Wood:2009a}, \opencite{Davis:2009} and \opencite{Davies:2009}) in the heliospheric imager field-of-view. 

\inlinecite{Lugaz:2009c} and \inlinecite{THoward:2009a} have proposed a different way to derive heliocentric distances from elongation angles. In these works, the authors assume that the CME can be modeled as a circular front whose center propagates on a fixed radial trajectory and which is anchored at the Sun. It is further assumed that the measured elongation angle simply corresponds to the angle between the Sun-spaceraft line and the line-of-sight tangent to this circular front. Here the direction, $\beta_\mathrm{HM}$, corresponds to the direction of the nose of the CME, i.e. the direction of the point on the circular front with the largest heliocentric distance. As derived in \inlinecite{Lugaz:2009c}:
$$
R_\mathrm{HM} = 2 d_\mathrm{ST}\frac{\sin \alpha}{1 + \sin(\alpha + \beta_\mathrm{HM})}.
$$
This relation has been referred to as the harmonic mean (HM) approximation, because the distance corresponds to the harmonic mean of the distance calculated with Fixed-$\Phi$ and Point-P approximations. It can be also inverted, assuming a constant velocity (CV), as:
\begin{eqnarray}
\alpha &= & \arctan \left (\frac{ V_\mathrm{HMCV}  t \sin\beta_\mathrm{HMCV} }{2 d_\mathrm{ST} - V_\mathrm{HMCV} t \cos\beta_\mathrm{HMCV}} \right) +  \label{eq:TEHM} \\
& & \arcsin \left (\frac{ V_\mathrm{HMCV} t  }{\sqrt{\left(2 d_\mathrm{ST} - V_\mathrm{HMCV} t \cos\beta_\mathrm{HMCV} \right)^2 + \left(V_\mathrm{HMCV}  t \sin\beta_\mathrm{HMCV} \right)^2}} \right). \nonumber
\end{eqnarray}

A measured profile of elongation vs. time can also be fitted to profiles of calculated elongations and the harmonic mean constant velocity (HMCV) procedure gives different estimates of the average speed and direction of the transient.

\subsection{Fitting Procedure}
In this section, we give a quick overview of the fitting procedure for the two methods just described. A more detailed explanation can be found in \inlinecite{Rouillard:2010} and \inlinecite{Davis:2010}. First, a time-elongation map (J-map) is produced from a time sequence of running difference HI images. The procedure to produce J-maps is explained in details in \inlinecite{Sheeley:1999} and \inlinecite{Davies:2009} and we refer interested readers to these works. From a J-map, a number of sample points, $N$ (typically 30 to 50), following the CME track are selected manually by an operator. These time series of time-elongation data are plotted on the Rutherford Appleton Laboratory website \footnote{\url{http://www.sstd.rl.ac.uk/stereo/HIEventList.html}} and they were provided to us by C.~Davis. The time-elongation data are the starting point of the current study. The typical error associated with the manual selection of the sample points has recently been addressed in \inlinecite{Williams:2009}. They estimated the error in the direction to be typically 2-5$^\circ$ for CMEs observed up to 45$^\circ$ elongation and beyond.

\begin{figure*}[t*]
\begin{center}
{\includegraphics*[width=6.cm]{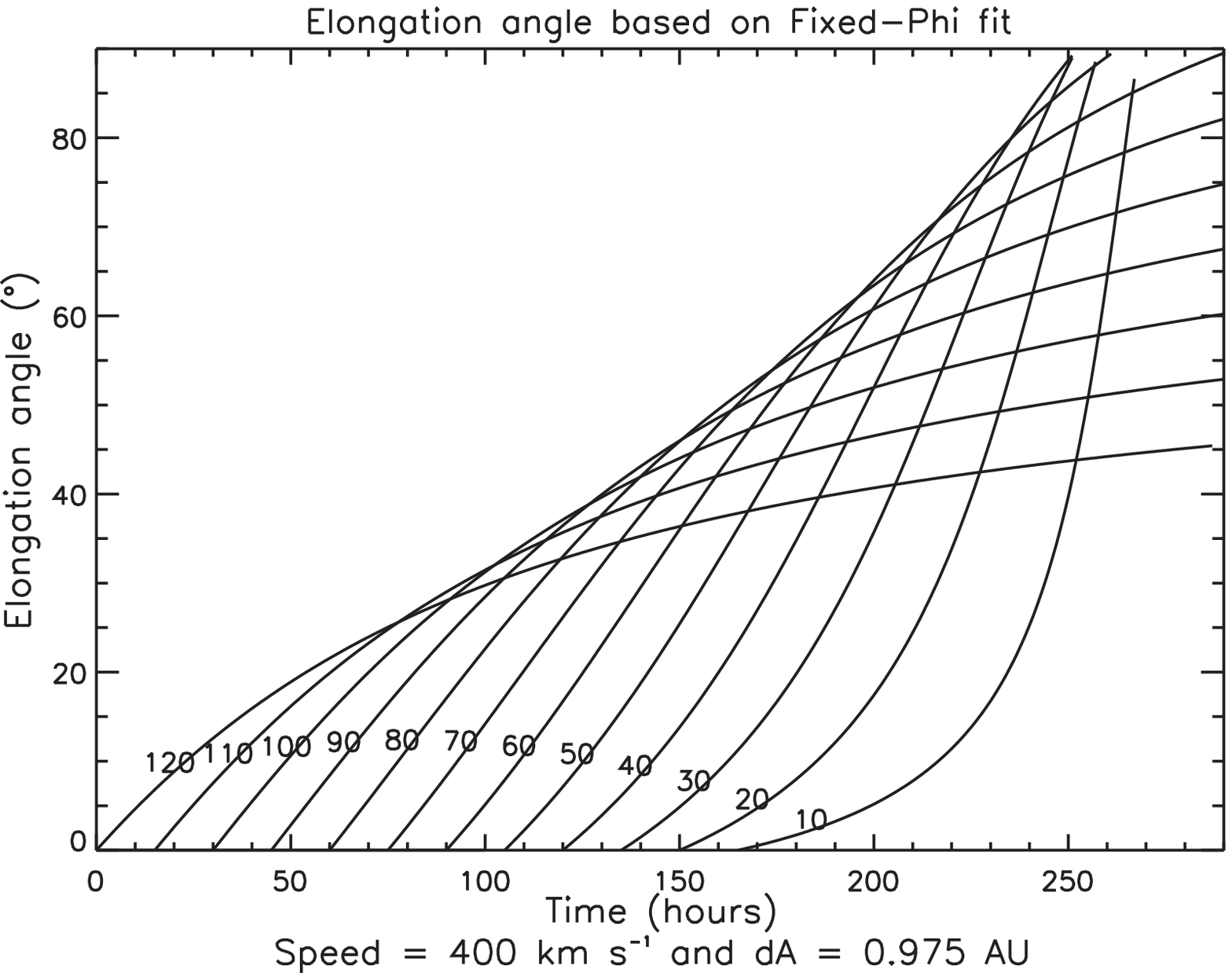}}
{\includegraphics*[width=6.cm]{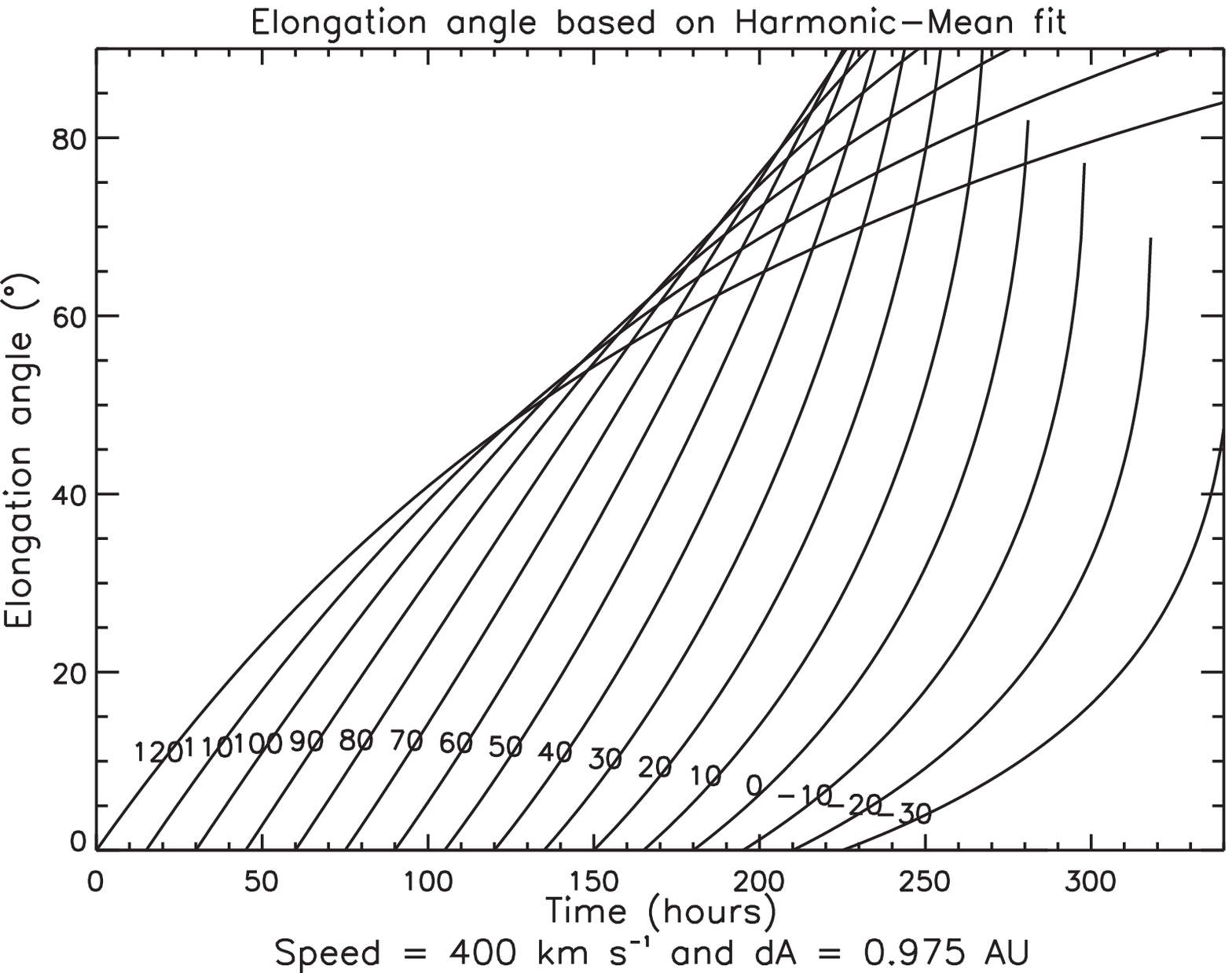}}
\caption{Time-elongation profile for a 400~km~s$^{-1}$ transient propagating with different directions with respect to the observing spacecraft assuming the position is given by the Fixed-$\phi$ approximation (left) and the Harmonic Mean approximation (right).}
\end{center}
\end{figure*}


A time series is fitted with the theoretical formulae given by Equations~(1) and (2) in the following manner. For a given value of the velocity, V,  and the CME direction, $\beta$, we calculate the standard error $\sigma$ (standard deviation of the residue) between the observed profile and the theoretical profile as follow:
\begin{equation}
\sigma^2 = \frac{1}{N} \sum_{k=1}^N \left( \alpha_{\mathrm{observed}}(t_k) - \alpha_{\mathrm{theoretical}}(t_k)\right)^2, 
\end{equation}
where $\alpha_{\mathrm{observed}}$ is the observed elongation angle at time $t_k$ and $\alpha_{\mathrm{theoretical}}$ is the elongation angle calculated with Equation~(1) or (2) for the same time. $t_k$ is the sample time of the $k^{\mathrm{th}}$ point on the J-map track. We only use elongation measurements in the HI fields-of-view, because the speed of the CMEs in the coronagraphic fields-of-view is usually not yet constant. Therefore $t_1$, the sample time corresponding to the first point on the J-map track is not obtained from measurements only. Solar measurements, such as flare time or first appearance in COR-1 field-of-view, can not be used to determine this time, since it typically corresponds to an elongation of $4-5^\circ$. For each value of V and $\beta$, $t_1$ is obtained by assuming a constant speed and by solving $\alpha_{\mathrm{theoretical}}(t_1) = \alpha_{\mathrm{observed}}(t_1)$. It yields a different value of $t_1$ for each of the two fitting methods and for each value of V and $\beta$.

The procedure is repeated for values of the speed between 100 and 1000 km~s$^{-1}$ by 1~km~s$^{-1}$ increment and for values of the direction between $-30^\circ$ and 120$^\circ$ with 1$^\circ$ increment. This way, we obtain an ``error map'' giving the value of $\sigma$ for all possible combinations of V and $\beta$. The best-fit values of (V, $\beta$) is that for which $\sigma$ is minimum. We give the uncertainties in the fitting quantities corresponding to the value of (V, $\beta$) for which $\sigma = 2\sigma_\mathrm{min}$, corresponding to a 95$\%$ certainty.  A theoretical example is given in the following section, and an example based on actual data in Section~\ref{comp}.

\begin{figure*}[t*]
\begin{center}
{\includegraphics*[width=5.4cm]{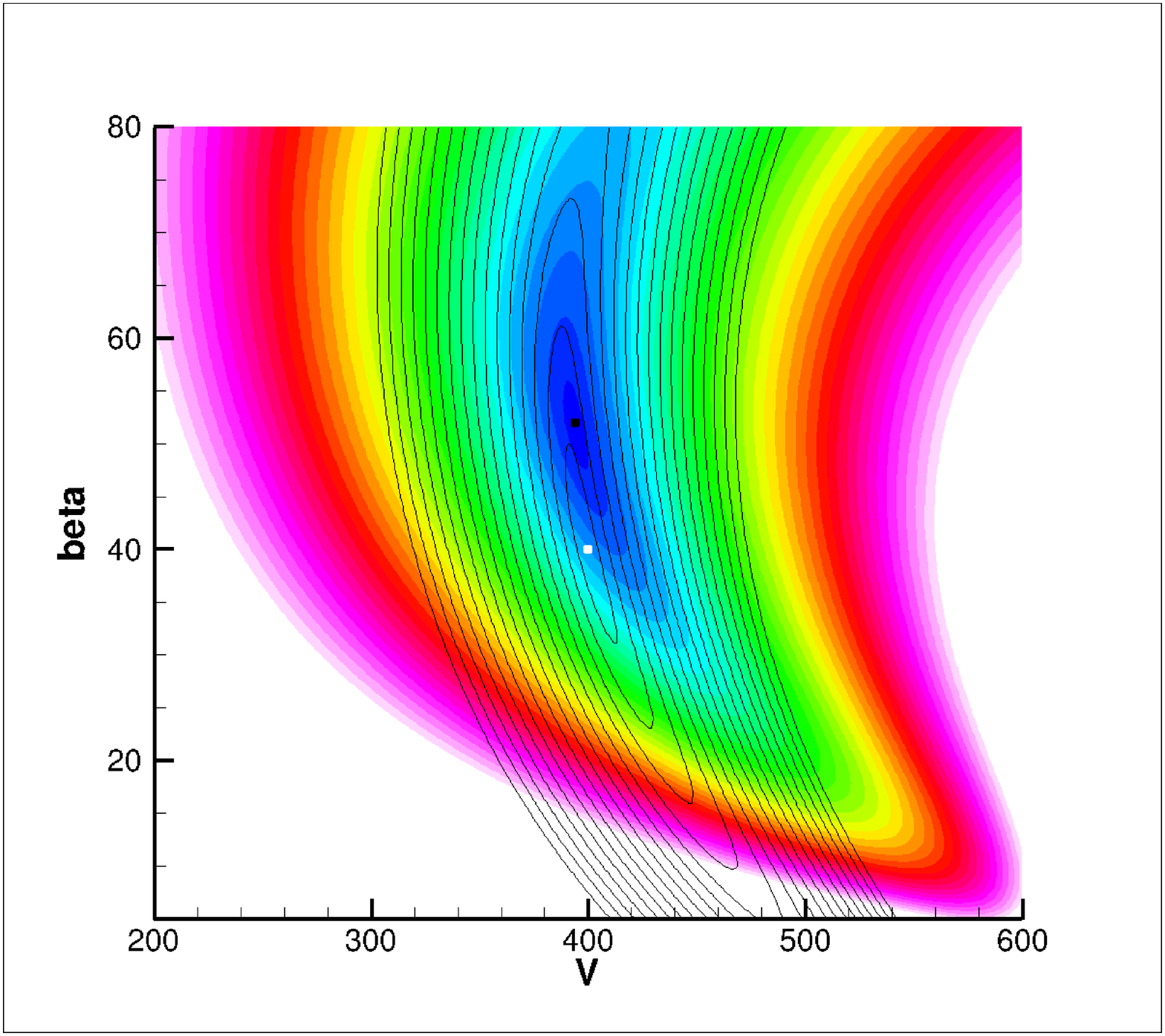}}
{\includegraphics*[width=6.6cm]{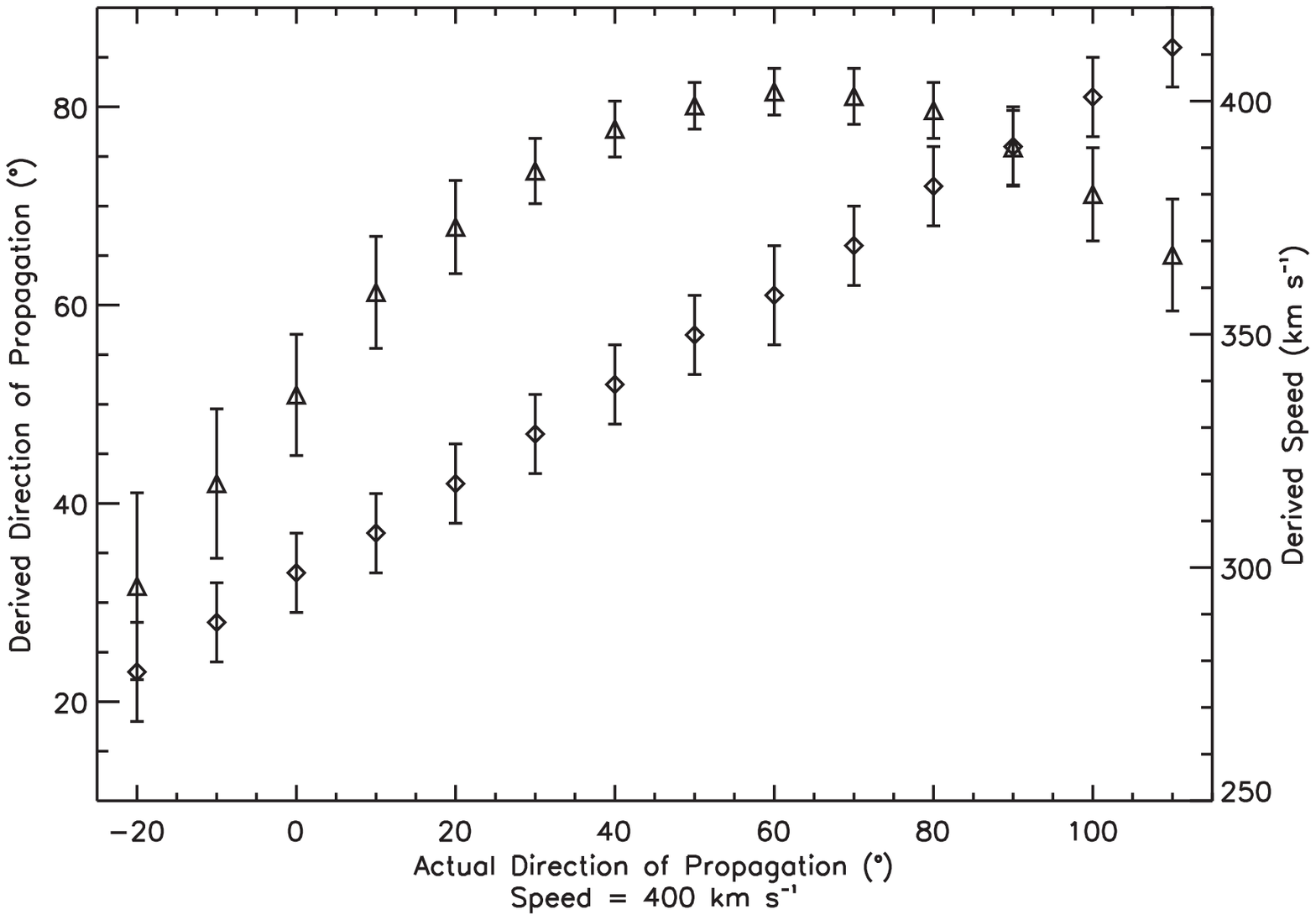}}
\caption{Left: Error in the fit for an analytical profile given by Equation~(2) (HMCV) with a speed of 400~km~s$^{-1}$ and a direction of 40$^\circ$. Color contours are the fitting error for the fit with Equation~(1) (best fit marked by the black circle, blue is small error, white large) and line contours are the fitting error for the fit with Equation~(2) (best fit marked by the white circle). Right: Best-fit direction (black diamonds) and velocity (red triangles) for a profile given by Equation~(2) (HMCV) for a 400~km~s$^{-1}$ CME observed from different directions and analyzed with Equation~(1) (F$\Phi$CV).}
\end{center}
\end{figure*}

\subsection{Theoretical Comparison}
In this section, we compare theoretically the two fitting methods. Figure~1 shows elongation vs time profiles given by Equations~(1) and (2) for different directions of propagation. All cases correspond to a CME with a speed of 400 km~s$^{-1}$ observed by STEREO-A. A similar figure was shown in \inlinecite{Webb:2009} but the authors never used this as a fitting procedure to derive the direction of propagation of CMEs.

As can be seen in this Figure, for a given direction, the new fitting method (HMCV) predicts that a CME exhibits less apparent deceleration/acceleration than as predicted by the F$\Phi$CV fitting method of \inlinecite{Sheeley:1999}. The reason is that, as a wide CME propagates close to the solar limb, an heliospheric imager observes more and more of the flank of the CME. It results in a apparent deflection towards the spacecraft of the observed structure and in a reduction of the ``deceleration'' due to geometrical effects. Conversely, for CMEs propagating towards the observing spacecraft, more and more of the CME nose is observed, resulting in less apparent acceleration.

Next, we consider how a time-elongation profile given by Equation~(\ref{eq:TEHM}) is fitted with profiles from Equation~(\ref{eq:TEFF}) using the procedure explained in the previous section. This gives an idea of the theoretical error associated with the fact that the fitting method of \inlinecite{Sheeley:1999} assumes that the heliospheric imager tracks a single plasma element and not a front.

In the left panel of Figure~2, we show an example of an error map corresponding to a CME front propagating 40$^\circ$ away from the Sun-spacecraft line with a speed of 400 km~s$^{-1}$. A synthetic time-elongation profile for these values of $V$ and $\beta$ is generated using Equation~(2). Fitting this profile with the procedure of \inlinecite{Sheeley:1999} results in a best-fit speed $V_{\mathrm{F}\Phi CV} = 394 \pm 6$ km~s$^{-1}$ and a direction $\beta_{\mathrm{F}\Phi CV} = 52^\circ \pm 4^\circ$. The error in the velocity is negligible but the error in the direction of propagation is relatively large. It is about twice as much as the error associated with the manual selection of points \cite{Williams:2009}. This result holds true for other directions, as shown in the right panel of Figure~2. In fact, we found simple relations between the direction of propagation and speed based on the F$\Phi$CV fitting of \inlinecite{Sheeley:1999} compared to that based on the HMCV fit:

\begin{eqnarray}
\beta_{\mathrm{HMCV}}(^\circ) &=& 2.067 \beta_{\mathrm{F}\Phi CV} - 67.3 \\
\Delta V (\%) &=& 0.0168 \alpha_{\mathrm{F}\Phi CV}^2 - 2.12 \alpha_{\mathrm{F}\Phi CV} + 66.38, 
\end{eqnarray}
with $\Delta V = \left(V_\mathrm{HMCV}  - V_{\mathrm{F}\Phi CV} \right) / V_\mathrm{HMCV}$ and with correlations of  0.99985 and 0.9988, respectively. This is found independently of the CME speed. It shows that around 60$^\circ$, both fitting methods give the same direction. The F$\Phi$CV fitting gives large errors in the direction for synthetic profiles corresponding to a wide CME with a direction greater than $80^\circ$ or less than $40^\circ$ with respect to the Sun-spacecraft line.

\section{CME Direction of Propagation from Stereoscopic Observations} \label{stereo}

CMEs propagating between the STEREO-A and B spacecraft or wide CMEs propagating close to one of the STEREO spacecraft can be imaged to large elongation angles by the HIs onboard both STEREO spacecraft. Simultaneous measurements can be used to derive the CME direction of propagation for every pair of observations. Direct triangulation can be done under the assumption that both STEREO spacecraft observe the exact same plasma element \cite{Liu:2010}. Here, we derive a slightly different version of their formula, taking into account the difference in spacecraft heliocentric distances. The ratio of these distances is typically $d_B/d_A = 1.07$ and varies between about 1.04 and 1.14. Using the correct ratio results in a shift of the CME direction by as much as 30$^\circ$ at large distances (beyond 0.5 AU) compared to assuming the ratio equal to one.

The direction of propagation is given by
\begin{eqnarray}
\beta_{\mathrm{Triang}} &=& \arctan \left( \frac{P \sin(\alpha_A + \gamma_A) - \sin(\alpha_B + \gamma_B)}{P \cos(\alpha_A + \gamma_A) + \cos(\alpha_B + \gamma_B)} \right), \\ \label{eq:Triang}
& &  \mathrm{with} \nonumber \\
\bigskip
P & = & \frac{d_B \sin \alpha_B}{d_A  \sin \alpha_A},  \nonumber
\end{eqnarray}
where $\gamma_A$ and $\gamma_B$ are the separation between STEREO-A and B and Earth, respectively (both are defined as positive numbers).

Alternatively, one can use the model of \inlinecite{Lugaz:2010b}, which considers that the two STEREO spacecraft observe the tangent to a circular CME front anchored at the Sun. In this case, as shown in \inlinecite{Lugaz:2010b}, the direction of propagation is given by:

\begin{eqnarray}
\beta_{\mathrm{Tang}} &=& \beta_{\mathrm{Triang}} + \arcsin\left( \frac{P -1}{Q} \right),\\
& &  \mathrm{with} \nonumber \\
Q & = & \sqrt{P^2 + 2P \cos(\gamma_A + \gamma_B + \alpha_A + \alpha_B) + 1}, \nonumber
\end{eqnarray}
and $\beta_{\mathrm{Triang}}$ given by Equation~(6). 
This Equation is correct for $P \cos (\gamma_A + \alpha_A) + \cos(\gamma_B + \alpha_B) \ge 0$ (which is equivalent to $\alpha_A + \alpha_B + \gamma_A + \gamma_B \le \pi$ for $d_A = d_B$). When the inequality is no longer true, the correct solution is:
$$
\beta_{\mathrm{Tang}} = \beta_{\mathrm{Triang}} - \arcsin\left( \frac{P -1}{Q} \right).
$$
We refer to this method as the ``tangent-to-a-sphere'' method, because the elongation angle is understood as the direction of the tangent to the spherical CME front.

\section{Method Comparison Based on Real Data} \label{comp}
In this section, we compare these four methods on CMEs observed simultaneously by the two STEREO spacecraft in 2008 and 2009. This allows us to use the stereoscopic methods and also to compare the fitting results for each of the two spacecraft. We found about 15 CMEs with stereoscopic observations in these two years, 12 of which had good enough data available on the RAL website for the study. These are the 2008 April 26, June 2, July 7, August 30 and December 12 (two tracks: front and back) CMEs and the 2009 January 9, January 22, May 9, May 13, September 4, October 18 and November 21 CMEs. Here, we analyze in details the 9 January 2009 CME, before comparing statistically the different methods in the next section.

For an elongation vs. time profile, we judge one fitting method to be better than the other, when the best-fit error (as defined in Section~2.2) of this method is smaller than the best-fit error of the other method. It is also possible to compare the best-fit (V, $\beta$) obtained with the data from STEREO-A with that obtained from the data from STEREO-B for the same CME. When there are values of (V, $\beta$) within 1-$\sigma_\mathrm{min}$ of the best-fit value for STEREO-A and B data, we consider that there is good agreement between the two sets of data. A specific example is given in the following section.

\begin{figure*}[t*]
\begin{center}
{\includegraphics*[width=5cm]{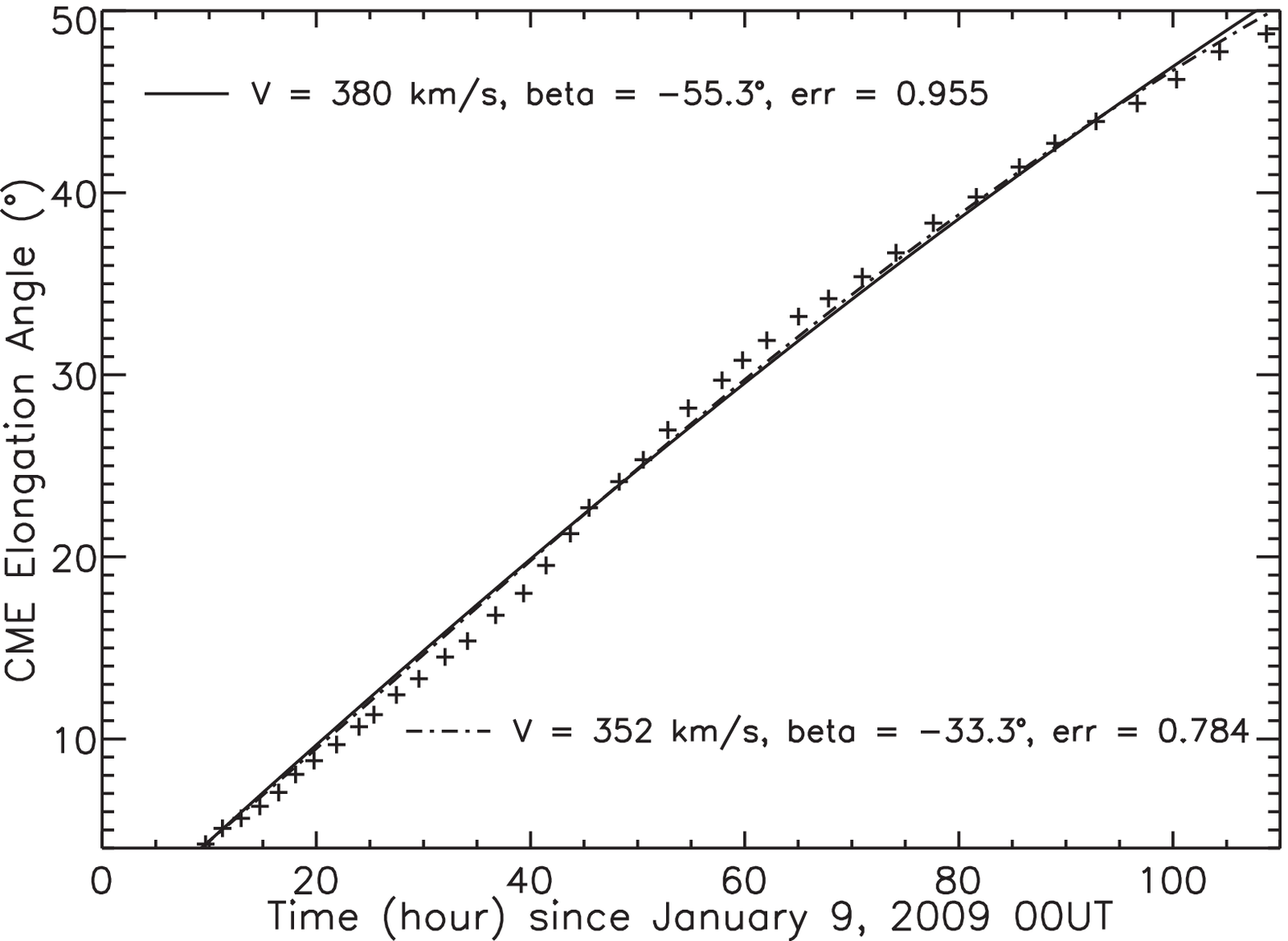}}\\
{\includegraphics*[width=5cm]{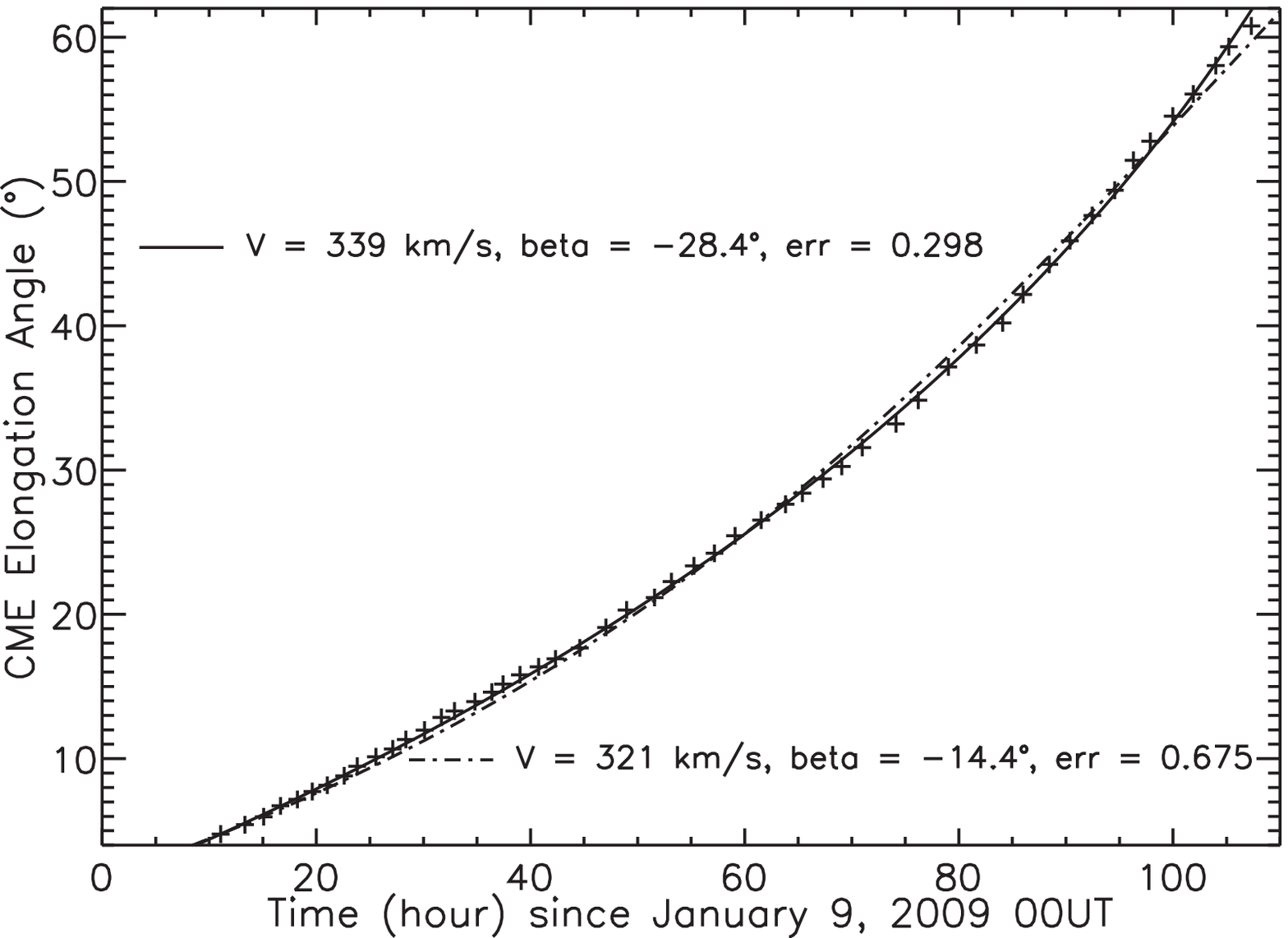}}\\
{\includegraphics*[width=5cm, angle=270]{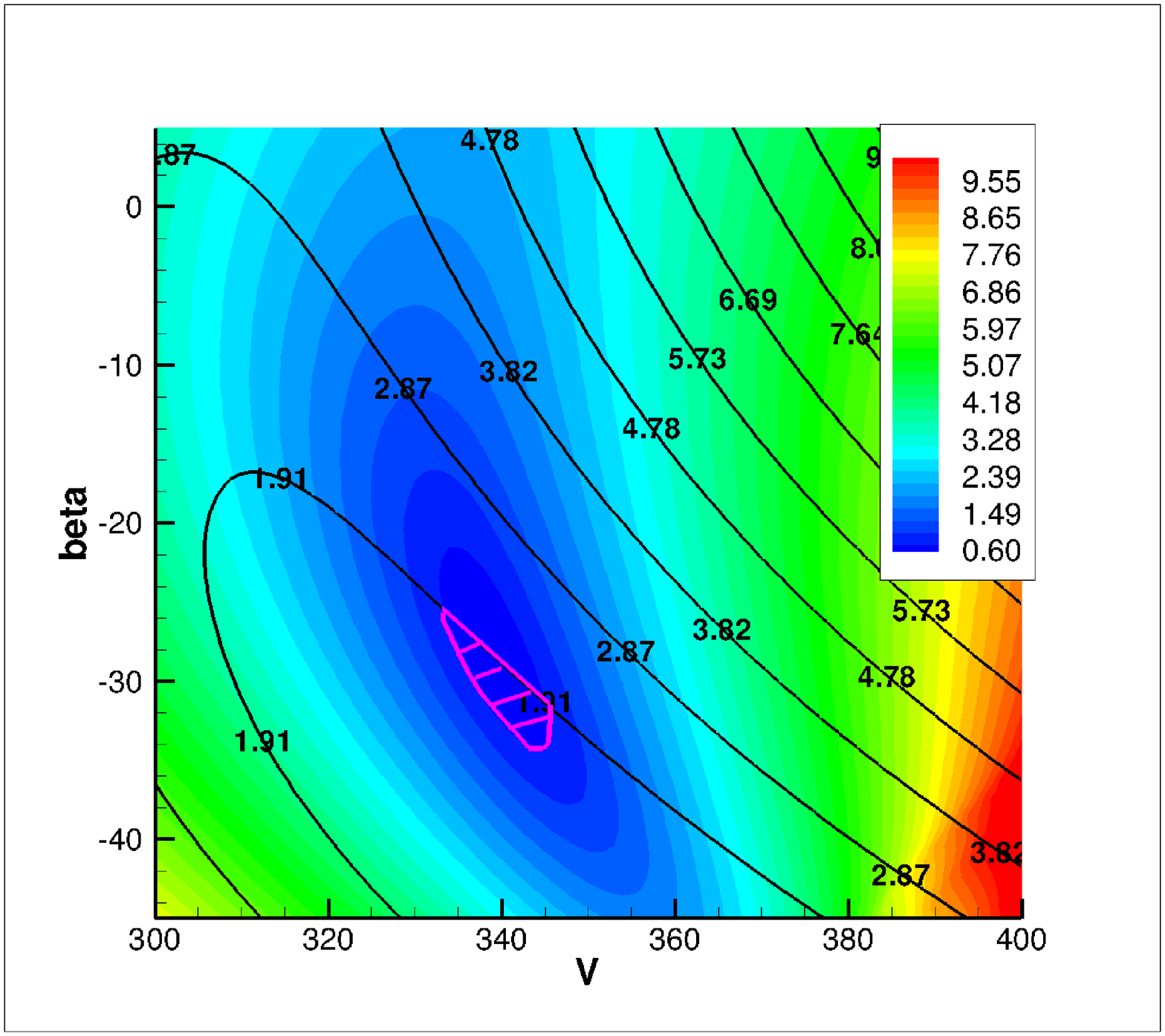}}
\caption{Analysis of the 9 January 2009 CME. {\it Top}: STEREO-A data (cross) and best-fit solution obtained with HMCV (solid line) and F$\Phi$CV (dash-dot line) fittings. {\it Middle}: Same but for STEREO-B data. {\it Bottom}: Combining the results of the HMCV fitting for both STEREO spacecraft. Line contours and color contours show the error for STEREO-A and B data, respectively (each contour is separated by $\sigma_\mathrm{min}$). The pink striped area corresponds to value of V and $\beta$ simultaneously within the 95$\%$ certainty region of both fits.}
\end{center}
\end{figure*}

\subsection{2009 January 9 CME}
In January 2009, the STEREO spacecraft were almost in quadrature (separation of 89$^\circ$). There was a weak CME on January 8 2009 around 20:00 UT detected in LASCO, COR2-A and COR2-B. It was observed in the HIs for about 100 hours starting on January 9. 
A magnetic cloud was detected by STEREO-B starting on January 13 at 05:00 UT and ending at 22:00 UT the same day, and no magnetic cloud was observed by STEREO-A or ACE. From the predicted arrival time at 1~AU, we believe the January 9 2009 CME is a good candidate for this magnetic cloud. The January 9 CME was tracked until about 50$^\circ$ by STEREO-A (39 datapoints) and until about 60$^\circ$ by STEREO-B (52 datapoints). 

We fit the data from the two spacecraft with the two fitting methods. The time-elongation profiles corresponding to the best-fit values of (V, $\beta$) are shown in Figure~3 for the two methods. The fitting with Equation~(1) (F$\Phi$CV) of the STEREO-A data yields a best-fit speed of 352 $\pm$ 61 km~s$^{-1}$  and a direction of $-33.3^\circ \pm 17^\circ$. The fitting with Equation~(2) (HMCV) of the same data yields a best-fit speed of 380 $\pm$ 75 km~s$^{-1}$  and a direction of $-55.3^\circ \pm 40^\circ$. The HMCV minimal error is about 22$\%$ larger than that the F$\Phi$CV minimal error. The uncertainty for both methods is relatively large because the time-elongation profile is close to linear. As expected from our theoretical analysis, for this CME observed almost at the limb by STEREO-A, the HMCV fitting yields a larger value of the direction with larger errors than the F$\Phi$CV fitting.

The F$\Phi$CV fitting of the STEREO-B data yields a best-fit speed of 321 $\pm$ 10 km~s$^{-1}$  with a direction of $-14.4^\circ \pm 7.5^\circ$ for B data. The HMCV fitting of the same data yields a best-fit speed of 339 $\pm$ 6 km~s$^{-1}$  with a direction of $-28.4^\circ \pm 6^\circ$ and the error is about 126$\%$ smaller than that from the F$\Phi$CV best-fit. Because the time-elongation profile shows a strong acceleration starting around $25^\circ$, the direction of the CME is much better constrained by the STEREO-B data than with the A data. STEREO-B observes the CME as a halo, and the result of a smaller direction based on the HMCV fit compared to the F$\Phi$CV fit is consistent with our theoretical analysis. While the difference of $14^\circ$ between the two directions may appear small, it is larger than the derived uncertainty. The much smaller error of the HMCV fit compared to the F$\Phi$CV fit is a strong indication that STEREO-B did not observe the same part of the front at all time as assumed by the F$\Phi$CV fitting.

\begin{figure*}[t*]
\begin{center}
{\includegraphics*[width=6cm]{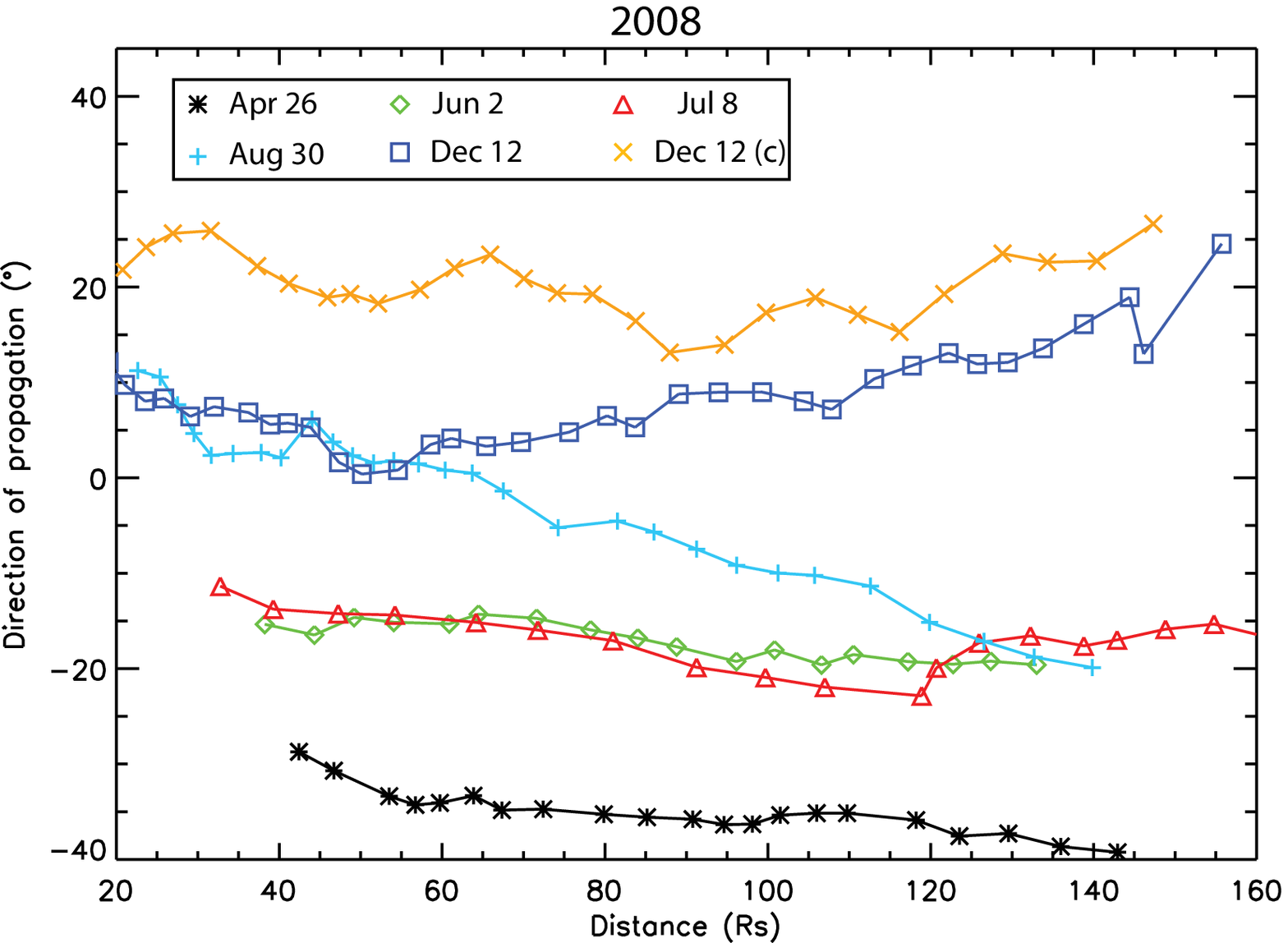}}
{\includegraphics*[width=6cm]{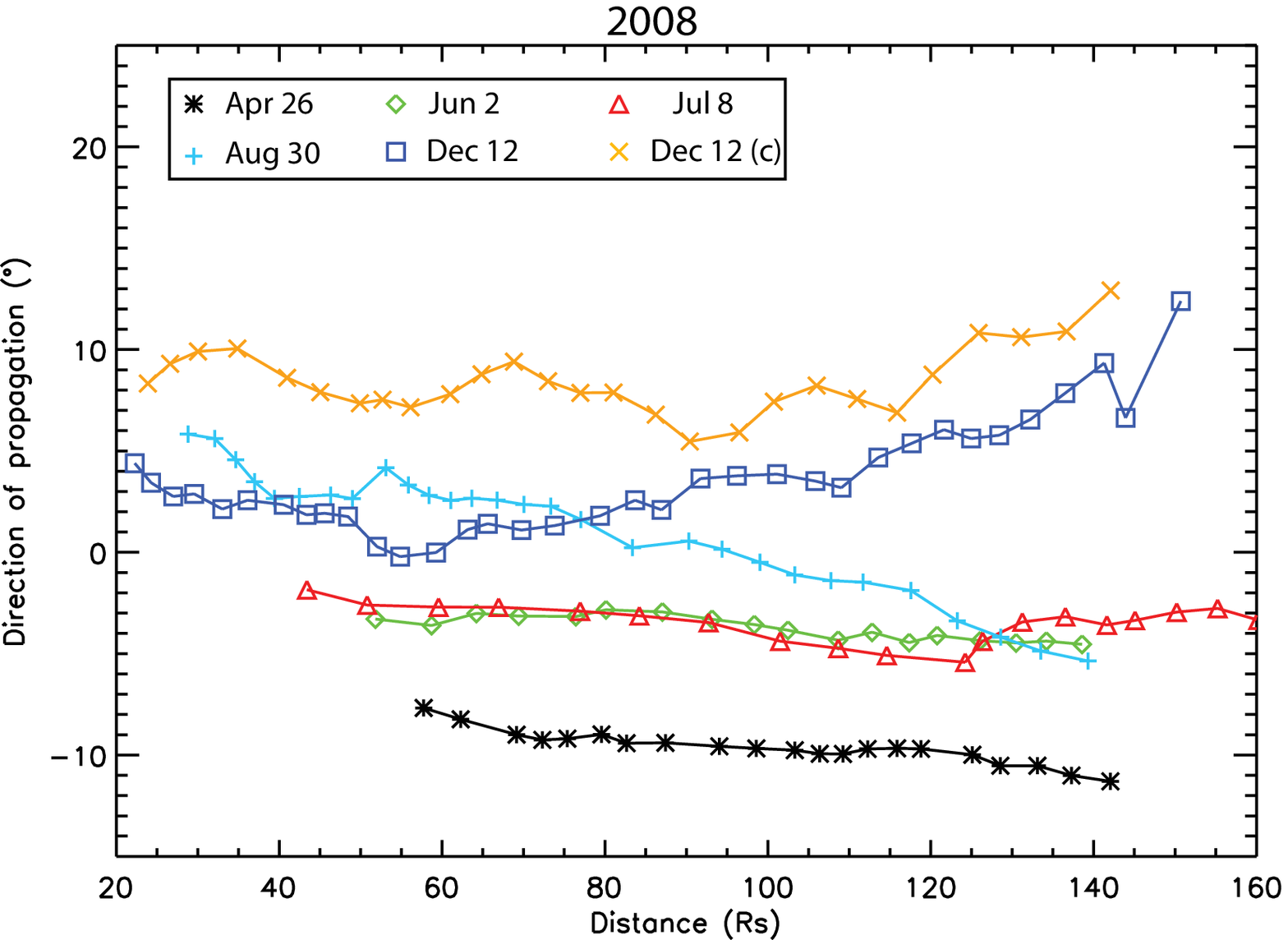}}\\
{\includegraphics*[width=6cm]{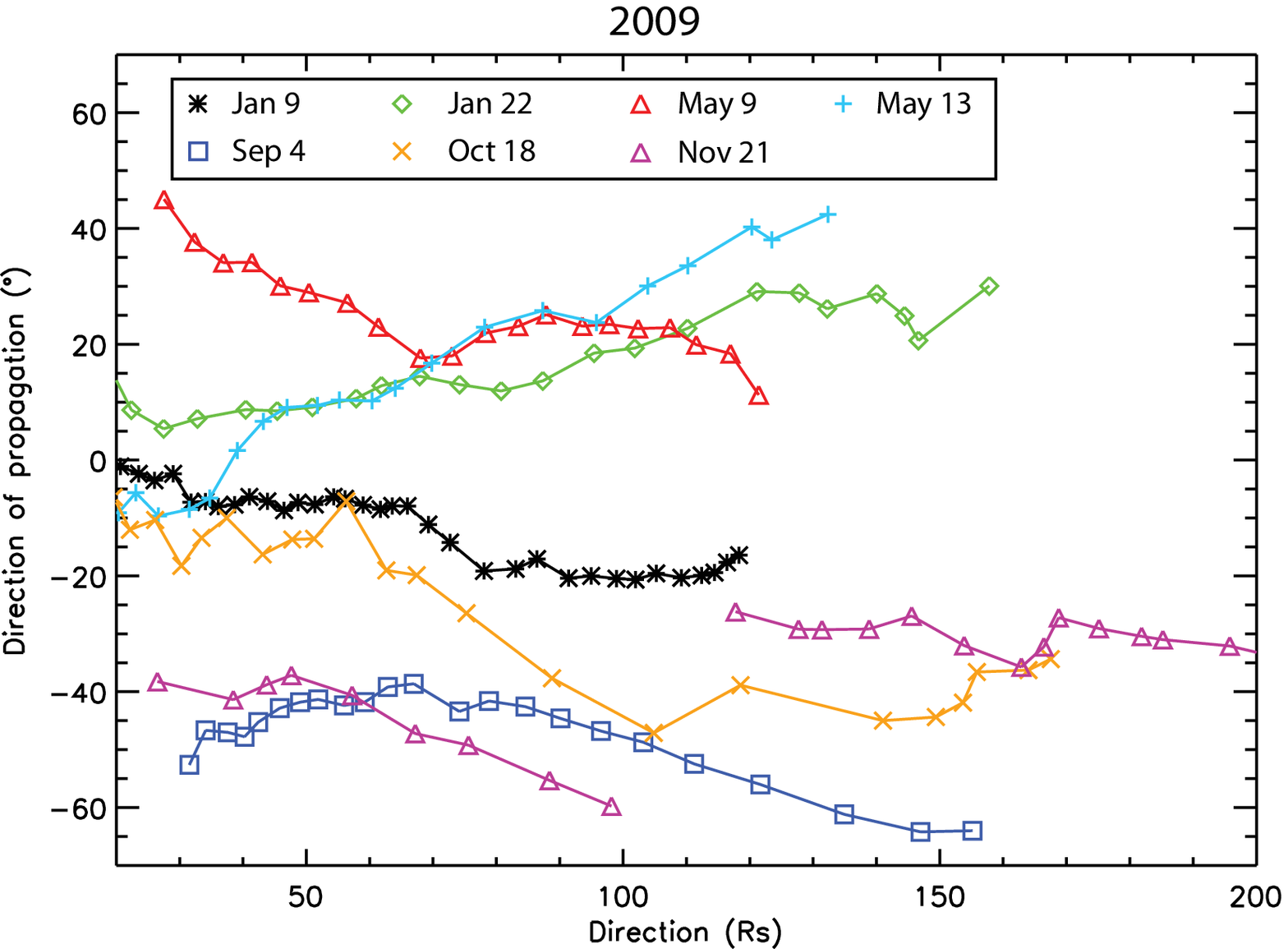}}
{\includegraphics*[width=6cm]{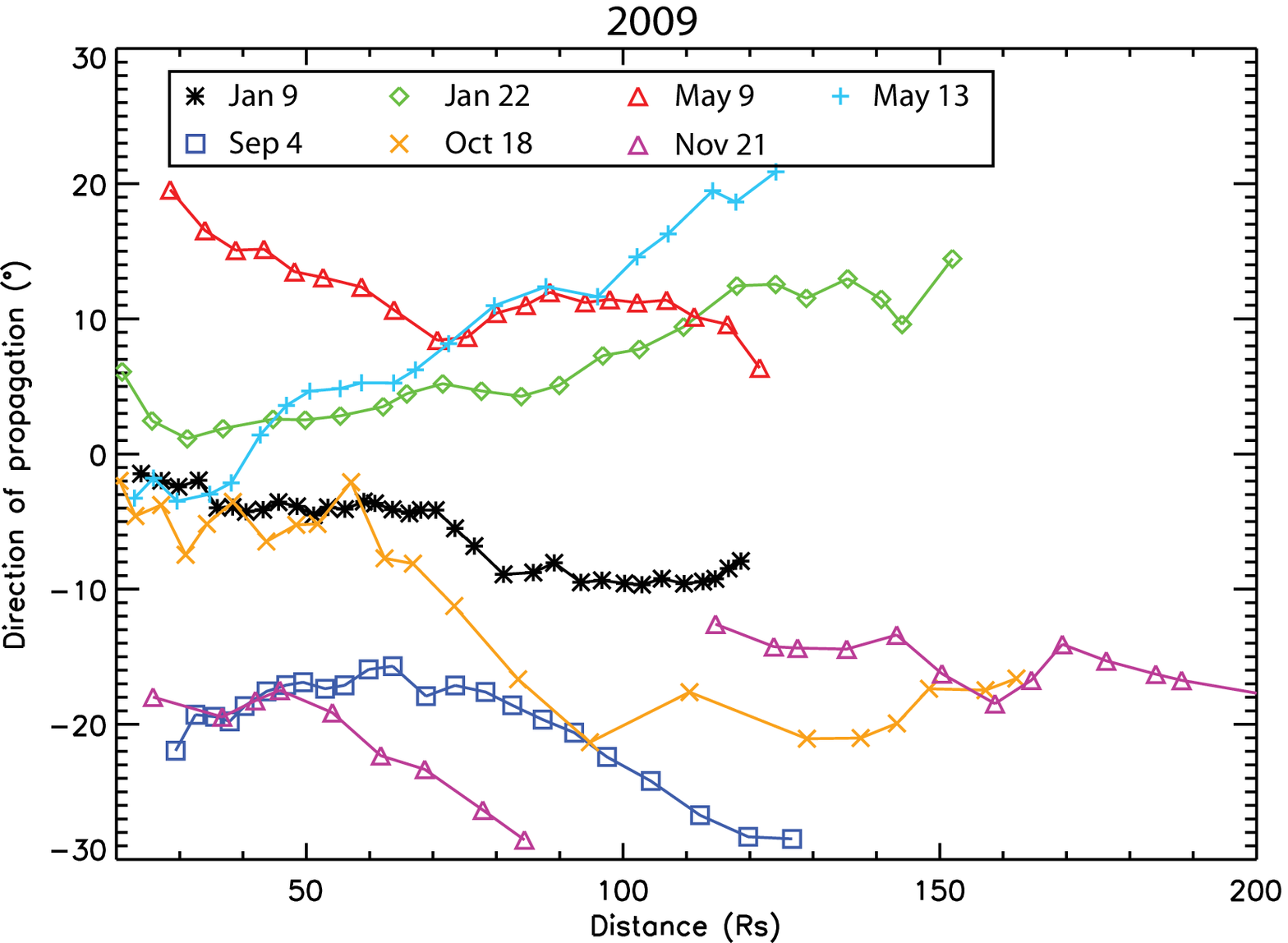}}\\
\caption{Direction of propagation with respect to the Sun-Earth line of the 13 CME tracks from Table 1 with the tangent-to-a-sphere method (left)  and the triangulation method (right). Note the different axis scales.}
\end{center}
\end{figure*}

We can combine the results of the two spacecraft by looking for values of (V,$\beta$) within 1-$\sigma_\mathrm{min}$ of the best values for STEREO-A and B data simultaneously. This is illustrated in the bottom panel of Figure~3 for the HMCV fit and it greatly reduces the uncertainties in the best-fit values. Combining the two set of observations, the best-fit value corresponds to a CME with a speed of 339 $\pm$ 6 km~s$^{-1}$ and a direction of $-30^\circ \pm 4.5^\circ$ (pink striped region). We perform the same procedure for the F$\Phi$CV fit (not shown here) and we find a best-fit value of the speed of 319 $\pm$ 8 km~s$^{-1}$ with a direction of $-18.5^\circ \pm 2.5^\circ$. If we assume that the magnetic cloud observed at STEREO-B on January 13 and not observed by ACE is associated with this ejection, and taking into account that STEREO-B was 46.6$^\circ$ away from Earth, the direction of $-30^\circ \pm 4.5^\circ$ from the HMCV fitting method appears more likely to be the correct one that $-18.5^\circ \pm 2.5^\circ$ from the F$\Phi$CV fitting.

Applying the stereoscopic methods to this CME, we find a direction of $-13 ^\circ \pm 7^\circ$ with the tangent-to-a-sphere method and $-6.5 ^\circ \pm 3^\circ$ for the triangulation technique limiting our analysis to distances between 30 and 120~$R_\odot$. The variation with distance of the direction of propagation from these two stereoscopic methods is shown in Figure~4. Both stereoscopic methods give results not in good agreement with the fitting methods and with a significant deflection towards the East. One interpretation is that the fitting methods give the direction of propagation at large heliospheric distances. If we assume a linear trend with distance, triangulation predicts a direction of about $-15^\circ$ at 1~AU and the tangent-to-a-sphere method predicts about $-30^\circ$. Based on these values, there is relatively good agreement between the two methods based on the Fixed-$\phi$ approximation (F$\Phi$CV and triangulation) and also good agreement between the two methods based on the harmonic mean approximation (HMCV and tangent-to-a-sphere). 

\begin{table*}[tb]
\centering
\begin{tabular}{cccccccc}
\hline
\hline
CME &  F$\Phi$CV A & F$\Phi$CV B & HMCV  A &  HMCV B & Triang. & Tangent & {\it in situ}\\
\hline
{\bf 2008}\\
Apr. 26 & $-33.5^\circ~\pm$18 & 2.1$^\circ~\pm$ 7 & ${\bf -34.5}^\circ \pm~{\bf 30}$ & ${\bf -19.9}^\circ~\pm$ {\bf 10} & $-9^\circ \pm 4$ & $-34^\circ \pm 5$  & B ($-24^\circ$)\\
June 2 & ${\bf -24.2}^\circ \pm$ {\bf 6}& $20.9^\circ \pm$ 12& $-34.2^\circ \pm~8$ & $-0.1^\circ \pm~25$ & $-3.5^\circ \pm$ 1 & $-16.5^\circ \pm$ 3 & B ($-25^\circ$) \\
July 8 & $-24.6^\circ \pm$ 7& ${\bf 2.6}^\circ \pm$ {\bf 11}& ${\bf -34.6}^\circ \pm~8$ & $-25.4^\circ \pm~21$& $ -3.5^\circ \pm$ 2 & $-17^\circ \pm$ 6  & none \\
Aug. 30 & ${\bf -0.4}^\circ \pm$ {\bf 8} & {\bf 19.2}$^\circ \pm$ {\bf 11} & $-19.4^\circ \pm$ 25 & 12.2$^\circ \pm$ 27 & 4$^\circ \pm$ 5 & 5$^\circ \pm$ 14 & none \\
Dec. 12 &  $-14.7^\circ \pm~13$ & 12.6$^\circ \pm~7$ &  ${\bf -14.7}^\circ \pm$ {\bf 19} & {\bf 18.6}$^\circ \pm$ {\bf 11} & 3$^\circ \pm$ 4 & 8$^\circ \pm$ 10 & ACE \\
Dec. 12 (b) & {\bf 8.3}$^\circ \pm$ {\bf 5} & -1.5$^\circ \pm$ 7 & 25.3$^\circ \pm$ 13 & ${\bf -14.5}^\circ \pm$ {\bf 11}& 8.5$^\circ \pm$ 3 & 20$^\circ \pm$ 6 & ACE\\
\hline
{\bf 2009}\\
Jan. 9 & ${\bf -33.3}^\circ \pm$ {\bf 17} & $-14.4^\circ \pm$ 8 & $-55.3^\circ \pm$ 40 & ${\bf -28.4}^\circ \pm$~{\bf 6} & $-6^\circ \pm$ 4 & $-11.5^\circ \pm$ 10 & B ($-46^\circ$)\\
Jan. 22 & $-10.1^\circ \pm$ 19 & {\bf 22}$^\circ \pm$ {\bf 8} & ${\bf -4.1}^\circ \pm$ {\bf 30} & 34$^\circ \pm$ 20 & 7$^\circ \pm$ 6 & 17$^\circ \pm$ 12 & none\\
May 9 & 10.4$^\circ \pm$ 6 & {\bf 8.1}$^\circ \pm$ {\bf 12} & {\bf 18.4}$^\circ \pm$ {\bf 9} & 5.1$^\circ \pm$ 32 & 12$^\circ \pm$ 4 & 23$^\circ \pm$ 11 & none \\
May 13 & ${\bf -14.3}^\circ \pm$ {\bf 8} & {\bf 18.1}$^\circ \pm$ {\bf 10} & $-27.3^\circ \pm$ 16 & $23.1^\circ \pm$ 22 & 12.5$^\circ \pm$ 8 & 16$^\circ \pm$ 25  & none\\
Sep. 4 & 5.6$^\circ \pm$ 11 & ${\bf -19.5}^\circ \pm$ {\bf 4} &  ${\bf 15.6}^\circ \pm$ {\bf 16} & $-32.5^\circ \pm$ 13 & $-18^\circ \pm$ 4 & $-50^\circ \pm$ 11 & none \\
Oct. 18 & ${\bf -19.8}^\circ \pm$ {\bf 12} & $-16.1^\circ \pm$ 7 & $-43.8^\circ \pm$ 26 & ${\bf -23.1}^\circ \pm$ {\bf 11} & $-18^\circ \pm$ 8 & $-30^\circ \pm$ 17& B ($-60^\circ$)\\
Nov. 21 & $-18.7^\circ \pm$ 10 & ${\bf -22.6}^\circ \pm$ {\bf 6} & ${\bf -44.7}^\circ \pm$ {\bf 17} & $-29.6^\circ \pm$ 14 & $-15.5^\circ \pm$ 4 & $-30^\circ \pm$ 4  & unknown\\
\hline
\end{tabular}
\caption{Direction of propagation for the 13 CME tracks from the four methods: F$\Phi$CV fitting method of Sheeley et al. (1999) for STEREO-A and STEREO-A data, HMCV fitting method for  STEREO-A and STEREO-B data, triangulation and tangent to a sphere methods for columns 2 to 7, respectively. For the fitting methods, bold results indicate that this method has the lowest error result for this dataset. When there are {\it in situ} measurements, we indicate it in column 8 as well as the spacecraft angular separation with Earth. (b) refers to the back of the December 12 CME.}
\end{table*}

\subsection{Summary of the Analysis}
The full results of our analysis are shown in Table~1, where directions are given with respect to the Sun-Earth line (positive number is towards the West, i.e. in the direction of STEREO-A).  In this Table, the results of the fitting method with the smallest error are marked in bold (better fit).  In addition to the 9 January 2009 CME, we are able to match the results of both fitting methods for STEREO-A and B spacecraft for the 2009 October 18  and November 21 CMEs. The results for the back of the 12 December 2008 CME can be similarly matched for the F$\Phi$CV fitting method of \inlinecite{Sheeley:1999}, but not with the HMCV fitting method. The contrary is true the 8 July  2008 CME (can be combined with the HMCV method but not with the F$\Phi$CV method). 

\begin{figure*}[t*]
\begin{center}
{\includegraphics*[width=6cm]{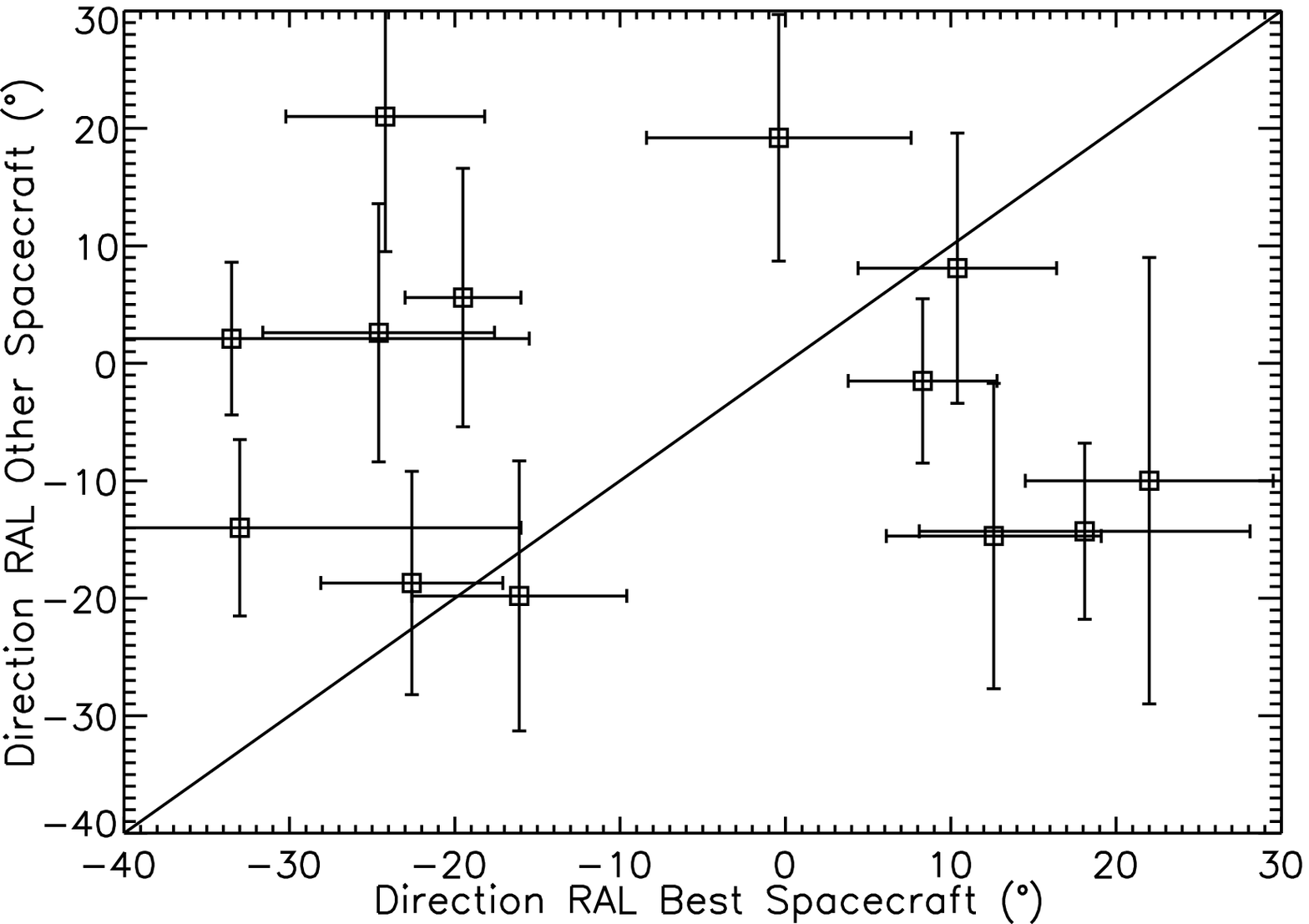}}
{\includegraphics*[width=6cm]{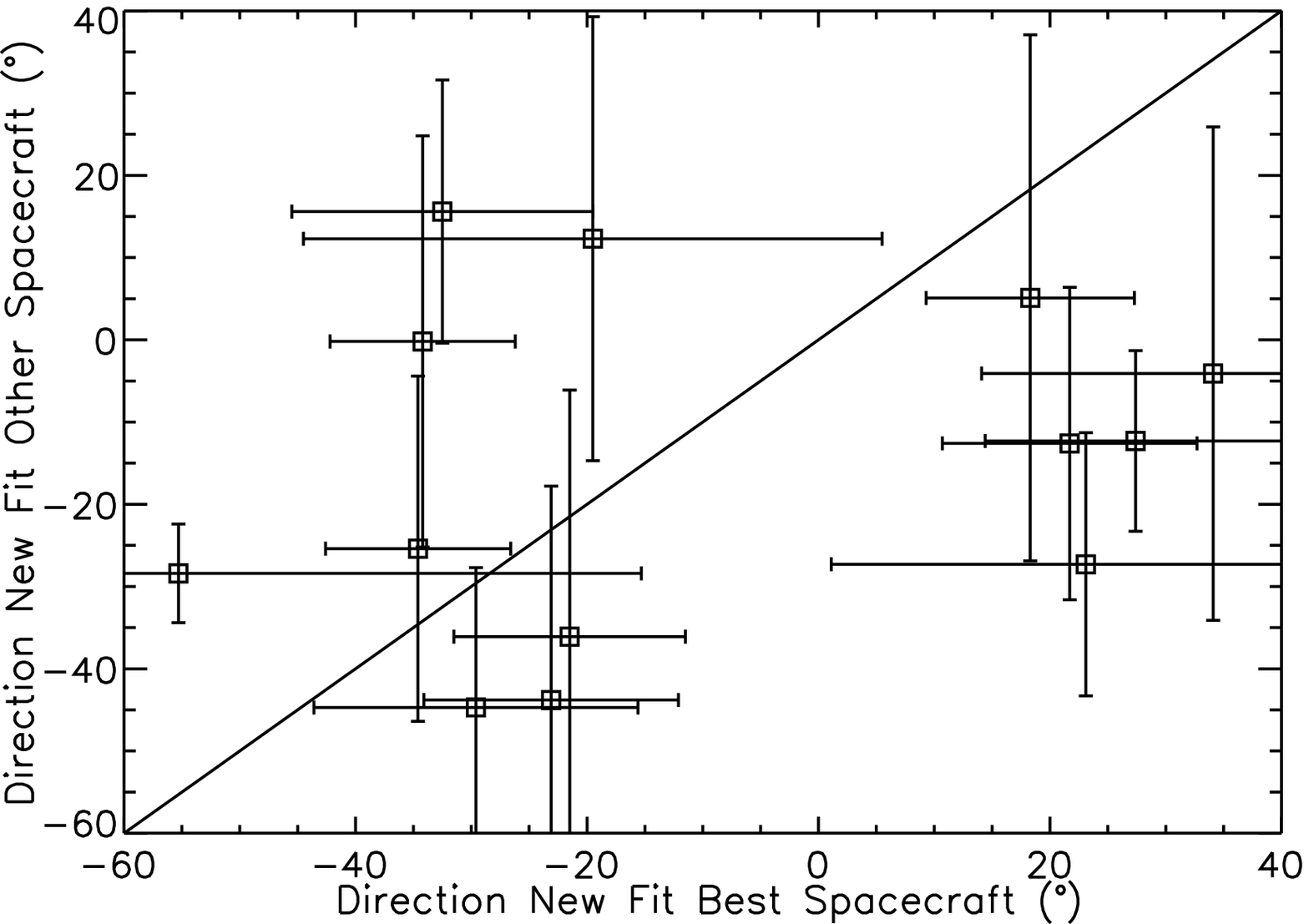}}
\caption{Comparison of the CME direction of propagation with respect to the Sun-Earth line obtained from fitting the observations by STEREO-A with that obtained from fitting the observations by STEREO-B with Equation~(1) (F$\Phi$CV, left) and Equation~(2) (HMCV, right).}
\end{center}
\end{figure*}

In addition to this analysis of CMEs observed by both STEREO spacecraft, we also analyze four CMEs which are observed away from the Thomson sphere: the 23 May 2008 CME observed by STEREO-B  with a direction of 13$^\circ$ with respect to the Sun-spacecraft line, the 16 August 2007 CME observed by STEREO-A with a direction of 32$^\circ$ away from the spacecraft, the December 4, 2007 CME observed by STEREO-B with a direction of 75$^\circ$ away from the spacecraft and the March 18 2008 CME observed by STEREO-A with a direction of 80$^\circ$ away from the spacecraft. The results of the analysis of these four CMEs observed by only one STEREO spacecraft are shown in Table~2.

\begin{table*}[b*]
\centering
\begin{tabular}{cccc}
\hline
\hline
CME &  F$\Phi$CV & HMCV  & Remarks\\
\hline
2007 Aug. 16 & $-17.6^\circ \pm 9$ & 8.4$^\circ \pm 18$   & Same fitting error\\
2007 Dec. 4 & $53.6^\circ \pm$ 7 & ${\bf 69.6}^\circ \pm$ {\bf 10}   & 71$^\circ$ from \inlinecite{Thernisien:2009}\\
2008 Mar. 18 & $-56.9^\circ \pm$ 12 & ${\bf -84.9}^\circ \pm$ {\bf 16}   & $-83^\circ$ from \inlinecite{Thernisien:2009}\\
2008 May 23 & $-11.7^\circ \pm$ 9 & ${\bf -37.7}^\circ \pm$ {\bf 4}   & Fit improved by 160$\%$\\
\hline
\end{tabular}
\caption{Direction of propagation for the four additional CMEs, selected because of the small or large direction of propagation with respect to the observing spacecraft.}
\end{table*}

\subsection{Comparing the Fitting Methods}
As shown in Table~1, the HMCV fitting method results in a better fit (as defined by a smaller fitting error) to the data about half the time (12/26 with one instance where there is no improvement). 
Comparing the direction of propagation obtained from STEREO-A data to the direction obtained from STEREO-B data, the HMCV fitting method yields results in slightly better agreement (see Figure~5), although statistically the cross-correlation between the direction obtained from STEREO-A and the direction obtained from STEREO-B only increases to 0.214 from $-0.125$. To detect any bias in the methods, it is enlightening to look at the direction of propagation with respect to the observing spacecraft. For this analysis, we treat all the observations as individual events and include the four CMEs from Table~2, resulting in 30 datapoints.
The results are shown in Figure~6, where better fits with the new method are shown with black squares and worst fits with red triangles. From this Figure, it appears clear that there is a systematic bias associated with one of the fitting methods, although it is slightly less pronounced than what we found in the theoretical analysis (Section 2.3). In fact, we can find a linear relation between the direction derived from the two methods with a cross-correlation of 0.966:
\begin{equation}
\beta_{\mathrm{HM}}(^\circ) = 1.835 \beta_{\mathrm{F}\Phi} - 45.  
\end{equation}
This relation is valid when we consider only the better fits or the entire dataset. This shows that the two methods result in CME directions within $\pm 15^\circ$ of each other for direction of $55^\circ \pm 18^\circ$, similar to what we found in the theoretical analysis. It is worth noting that out of these 30 CME tracks, the direction given by the F$\Phi$CV fitting and the direction given by the HMCV fitting are within each other 95$\%$ confidence range in all but ten cases. All these cases have a CME predicted direction of more than 70$^\circ$ or less than 20$^\circ$. For only one of these ten cases, the error of the F$\Phi$CV fitting is less than that of the HMCV fitting (plus one case where they are identical). 

\begin{figure*}[htb]
\begin{center}
{\includegraphics*[width=6cm]{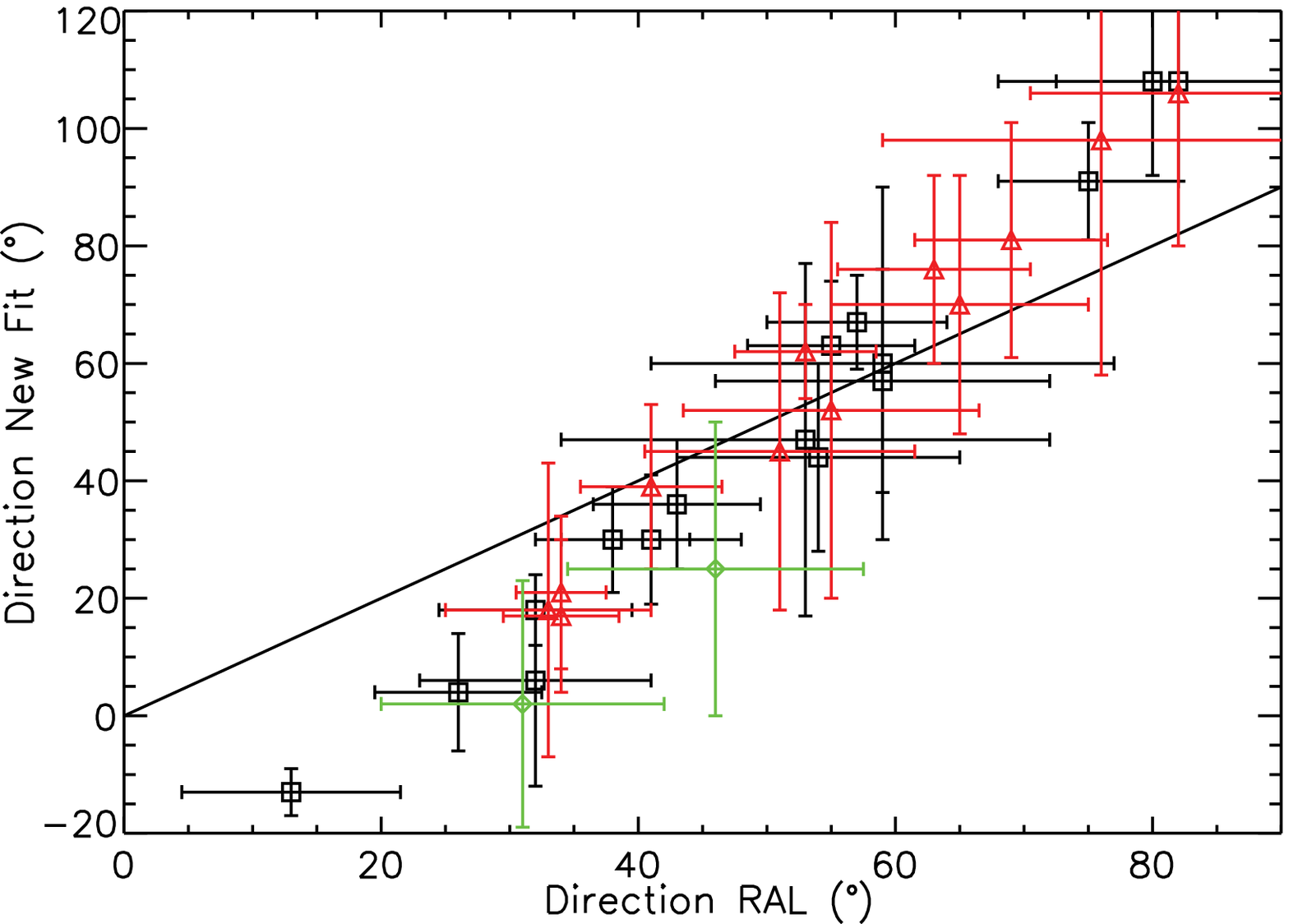}}
{\includegraphics*[width=6cm]{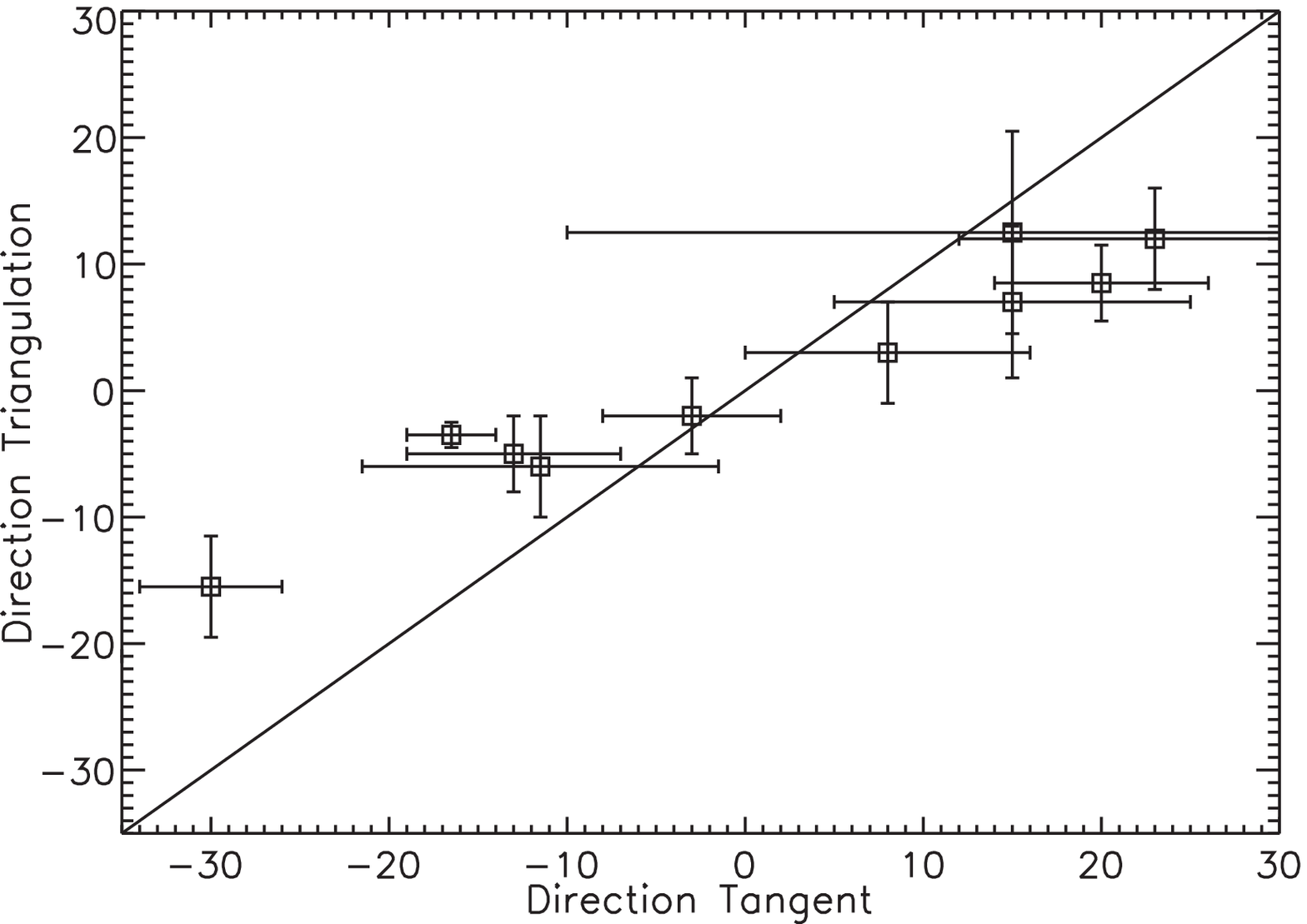}}
\caption{ Left: Comparison of the best-fit direction of propagation with respect to the Sun- spacecraft line obtained from the F$\Phi$CV fitting with that from the HMCV fitting. Red and black data points represent values for which the best-fit is obtained with the F$\Phi$CV and HMCV fittings, respectively and the green points are for cases when the best-fit error with the two methods is the same. Right: Comparison of the average direction obtained from the triangulation and from tangent-to-a-sphere method.}
\end{center}
\end{figure*}

Out of these 12 CMEs, at least five resulted in a magnetic cloud observed {\it in situ} by one of the spacecraft, providing some information of the real direction of propagation. Considering only the best observing spacecraft (the one which observes the CME to the largest elongation angle), both methods would have successfully predicted a CME hit with two exceptions: the 12 December 2008 CME, for which the HMCV fitting method may have predicted a miss at ACE or a hit also at STEREO-B and the 18 October 2009, for which the F$\Phi$CV fitting method would have predicted a miss at STEREO-B. 

Five of the studied CMEs were also part of the study by \inlinecite{Thernisien:2009}. They are the 4 December 2007 CME with a direction of $71^\circ$, 2008 March 18, April 26, June 2 and July 8 with direction of $-83^\circ$, $-21^\circ, -37^\circ$ and $-23^\circ$, respectively. For three of them, the direction obtained from the HMCV fitting method is closer to this direction from COR data than is the direction from the F$\Phi$CV fitting method. For the two other (April 26 and July 8), it is comparable or better if one uses the best-fit value from the spacecraft observing the CME as a halo. In fact, it is clear from our study that the direction is best constrained when it is less than 35$^\circ$ or more than 75$^\circ$, but when the CME can be tracked over a long-enough elongation range. Taking into account our theoretical analysis, the comparison of the direction with {\it in situ} measurements and with the procedure of \inlinecite{Thernisien:2009}, we conclude that the fitting method of \inlinecite{Sheeley:1999} has a systematic bias for CMEs propagating beyond 60$^\circ \pm 20^\circ$ away from the Sun-spacecraft line.

\subsection{Comparing the Stereoscopic Methods and CME Heliospheric Deflection}

Figure~6 shows the comparison of the two stereoscopic methods with each other. It also appears that there is a systematic bias, the triangulation method only yields directions of propagation very close to the Sun-Earth line. In fact, the two terms in Equation~(7) are almost always of the same order so that $\beta_\mathrm{Tang} \sim 2 \times \beta_{\mathrm{Triang}}$. It is less clear which of the two methods is systematically biased. Based on observations at 1~AU, triangulation works best for these CMEs propagating close to the Sun-Earth line such as the 12 December 2008 CME and the 4 September 2009 CME. It is also, in general, much less noisy than the tangent-to-a-sphere method. However, it gives inconsistent results for CMEs observed as a halo by one of the spacecraft (26 April and 2 June 2008 and 9 January  2009). For these CMEs, the stereoscopic method of \inlinecite{Lugaz:2010b} or a more complex analysis \cite{Wood:2009b,THoward:2009b} must be used. This result is expected as triangulation requires both spacecraft to observe the same plasma element. However, it points to a strong limitation of the triangulation method: it can  only be used if the direction can be estimated to be close to 0$^\circ$ by another method. To use triangulation in the HI field-of-views, it is necessary but not sufficient that both STEREO spacecraft observe the CME to large elongation angles. 

Next, we study the assumption of constant heliospheric direction. We focus on the tangent-to-a-sphere method, since the derived directions and their variations are about a factor of two lower based on the direct triangulation. This provides a worst-case scenario. For all of the 2008 CMEs except the August 30 ejection (four CMEs), the results are consistent with a very small heliospheric deflection (less than 15$^\circ$~AU$^{-1}$ in absolute value). This also true for the 21 November 2009 CME, if we only consider the measurements in HI-2 field of view (from 0.5 to 1 AU). For 2 other CMEs (2009 May 9 and September 4), the heliospheric deflection is found to be about 30$^\circ$~AU$^{-1}$, while the other five CMEs are found to have larger heliospheric deflection or noisy data. This difference between CMEs observed in 2008 and those observed in 2009 may indicate that stereoscopic measurements can only be made for moderate spacecraft separations (less than 90$^\circ$).
Taking into account the measurement errors and the limitations of the stereoscopic methods, we believe that a deflection by less than 20$^\circ$ during the duration of the observations, is consistent with the assumption of radial propagation in the HI fields-of-view. This is the case for seven of the 12 studied CMEs.

\section{Discussions and Conclusions} \label{conclusions} 

In this article, we have examined some of the methods used to derive the direction of propagation of coronal mass ejections (CMEs) from observations by heliospheric imagers. We have focused on four methods: the fitting method of \inlinecite{Sheeley:1999}, the triangulation method of \inlinecite{Liu:2010}, the tangent-to-a-sphere method of \inlinecite{Lugaz:2010b} and a new fitting method derived in this article.

Because it is impossible to know the exact direction of propagation of a CME, we have focused our analysis of real data on ejections observed simultaneously by the two STEREO spacecraft. For these CMEs, it is possible to use the stereoscopic methods and also to compare the direction of propagation based on fitting methods for each of the instruments. Overall, we have found a very small or no correlation between the direction obtained from observations by one spacecraft and that obtained from observations by the other spacecraft, and this, both for the fitting method of \inlinecite{Sheeley:1999} and for the new method we proposed. However, for four out of 13 CME tracks, it is possible to combine the results of the fit based on the data from each spacecraft within the 95$\%$ certainty interval to find a well-defined direction. 

We have addressed two of the three possible sources of error for the fitting method of \inlinecite{Sheeley:1999} applied to CMEs: i) the assumption of radial propagation (absence of heliospheric deflection), and ii) the assumption that the CME is of negligible width (or that the same part of the CME is always observed). We addressed i) by using stereoscopic methods which can give at all time the CME direction of propagation instead of its average value.
Stereoscopic methods need to be better tested and validated but for about nine out 13 CMEs, we have found reasonable results with the method of \inlinecite{Lugaz:2010b} and for two CMEs with the triangulation method of \inlinecite{Liu:2010}. For the other two cases, the method of \inlinecite{Lugaz:2010b} is too noisy and triangulation results in a large underestimation of the CME direction of propagation. In seven of these good cases, we have found no sign of strong heliospheric deflection of CMEs, the total deflection being less than 20$^\circ$ in the HI field-of-view. This result validates the usage of fitting methods to determine CME direction, since these methods assume no heliospheric deflection.

We have proposed a new fitting method which takes into account the CME width. We have found that the assumption of negligible CME width yields worst fit for the method of \inlinecite{Sheeley:1999} compared to this new method for CMEs propagating outside of $60^\circ \pm 20^\circ$ from the Sun-spacecraft line. {\em We believe this new method should be used for halo and limb CMEs as seen by the observing spacecraft}.  However, both methods can be easily used for all observations and the one with the smallest residual error should be chosen to provide the CME estimated speed and direction on a case-by-case basis. In our analysis of 30 CME tracks, we have found that each fitting method gives a better fit in about half the cases. We have also found that, when a CME is observed to large elongation angles by both STEREO spacecraft, the CME direction is best determined using the data from the spacecraft which observed the CME as a halo. This is especially true with the new fitting method.

There is a third issue with the two fitting methods:  the assumption of constant heliospheric velocity. All the CMEs in our sample (with the exception of the 26 April 2008 with a speed of about 600--700 km~s$^{-1}$) have speed between 280 and 450~km~s$^{-1}$ and we believe this assumption is approximatively true. This issue is hard to address because of the lack of fast CMEs in the past three years. Numerical simulations combined with a proper treatment of the Thomson scattering \cite{Lugaz:2008b, Lugaz:2009b, Manchester:2008, Odstrcil:2009} 
are one of the only ways to address this question of the effect of CME deceleration and we plan to look into this in future research. 

 \begin{acks}
The research for this manuscript was supported by  NSF grant ATM-0819653 and NASA grants NNX-07AC13G and NNX-08AQ16G.
SoHO and STEREO are projects of international cooperation between ESA and NASA. N.~L. would like to thank C.~J.~Davis and J.~A.~Davies from STFC/RAL for providing the time-elongation data, and
C.~M{\"o}stl, I.~I.~Roussev and A.~Vourlidas for useful discussions and the anonymous reviewers for their help in improving the manuscript.
   The SECCHI data are produced by an international consortium of 
  Naval Research Laboratory, Lockheed
  Martin Solar and Astrophysics Lab, and NASA Goddard Space Flight
  Center (USA), Rutherford Appleton Laboratory, and University of
  Birmingham (UK), Max-Planck-Institut f{\"u}r Sonnensystemforschung
  (Germany), Centre Spatiale de Liege (Belgium), Institut d'Optique
  Th{\'e}orique et Appliqu{\'e}e, and Institut d'Astrophysique
  Spatiale (France).  SoHO is a project of international cooperation between ESA and NASA,  and the SOHO LASCO/EIT catalogs are  maintained by NASA, the Catholic University of America, and the US Naval Research Laboratory (NRL).
\end{acks}

\bibliographystyle{spr-mp-sola-cnd}

\end{article}

\end{document}